\documentclass[prl,showpacs,preprintnumbers,amsmath,amssymb,twocolumn]{revtex4}

\usepackage{graphicx}
\usepackage{dcolumn}
\usepackage{bm}
\usepackage{color}

\begin{document}


\title{Interlaced Dynamical Decoupling and Coherent Operation of a Singlet-Triplet Qubit}

\date{\today}

\author{C.~Barthel$^{1}$}
\author{J.~Medford$^{1}$}
\author{C.~M.~Marcus$^1$}
\author{M.~P.~Hanson$^2$}
\author{A.~C.~Gossard$^2$}
\affiliation{$^1$Department of Physics, Harvard University, Cambridge, Massachusetts 02138, USA\\
$^2$Materials Department, University of California, Santa Barbara, California 93106, USA
}

\begin{abstract}
We experimentally demonstrate coherence recovery of singlet-triplet superpositions by interlacing qubit rotations between Carr-Purcell (CP) echo sequences. We then compare performance of Hahn, CP, concatenated dynamical decoupling (CDD) and Uhrig dynamical decoupling (UDD) for singlet recovery. In the present case, where gate noise and drift combined with spatially varying hyperfine coupling contribute significantly to dephasing, and pulses have limited bandwidth, CP and CDD yield comparable results, with $T_2\sim 80~\mu$s.
\end{abstract}


\maketitle

The singlet ($S =  \left(\mid\uparrow\downarrow\rangle\,-\,\mid\downarrow\uparrow\rangle\right)/\sqrt{2}$) and $m=0$ triplet ($T_0 =  \left(\mid\uparrow\downarrow\rangle\,+\,\mid\downarrow\uparrow\rangle\right)/\sqrt{2}$) spin states of two electrons in a double quantum dot form a versatile qubit that is inherently protected from collective dephasing~\cite{Levy02} and allows sub-nanosecond  rotation around one axis via electrical control of the exchange interaction~\cite{petta05}. Rotations around a second axis, needed for universal control, can be induced by a Zeeman field difference between the two quantum dots, created by a proximal micromagnet \cite{Obata} or controlled Overhauser fields~\cite{Foletti09}. Fast measurement of a qubit state has been demonstrated for two-electron spin states~\cite{MeunierPRB06,Singleshotpaper,Dotsensorpaper}, and also for single electron spin states~\cite{elzerman04b,Amasha08}. 
 
Temporal fluctuations of Overhauser fields evolve slowly compared to gate operation times~\cite{ReillyCorr08}, allowing a simple one-pulse Hahn echo \cite{Hahnecho} to significantly extend qubit coherence~\cite{petta05}. By repeating $\pi$-pulses on shorter intervals, Carr-Purcell (CP)  sequences \cite{CP, CPisCPMG} can extend qubit coherence to impressively long times~\cite{BluhmT2}. Theoretical work and experiments in other systems suggests that, depending on the environmental noise spectrum, echo sequences more complex than CP may further increase coherence times ~\cite{Khodjasteh,WitzelCDD2007,Uhrig2007,Lee2008,Biercuk2009, Pasini2010}. To date, experimental demonstration of decoupling schemes for spin qubits have been limited to recovery of an initially prepared singlet state. For applications in quantum information processing, however, decoupling schemes must preserve an arbitrary qubit state~\cite{Lidar}. 
\begin{figure}[b]
\includegraphics{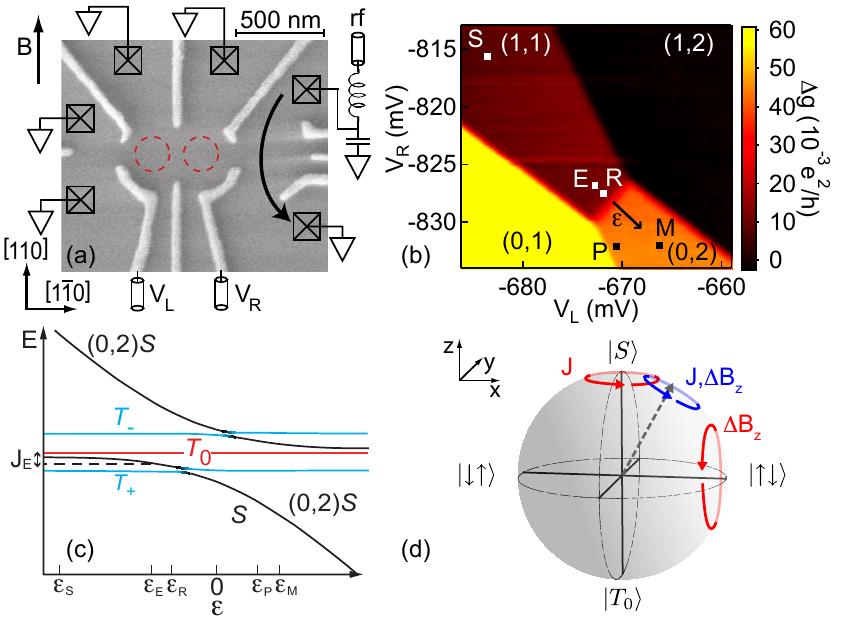}
\caption{\label{T2Fig1}~(Color online)~(a)~Micrograph of lithographically identical device with contacts, dot locations, and GaAs crystal axes indicated. Gate voltages, $V_{\rm{L}}$ and $V_{\rm{R}}$, set the energy of left and right dot. An rf-coupled sensor dot measures the double dot charge state.
(b) dc conductance change, $\Delta g$, with double dot charge state.  Markers indicate gate voltages  during experimental pulse sequences. Detuning, $\epsilon$, is controlled by $V_{\rm{L}}$ and $V_{\rm{R}}$ along the diagonal through points S and M. (c) Energy level diagram as function of detuning $\epsilon$, indicating locations of $\epsilon_{\rm{P}}$ where the (0,2) singlet is prepared, $\epsilon_{\rm{S}}$ of separated electrons for dephasing and $x$-rotations, $\epsilon_{\rm{E}}$ of exchange pulse and $\pi$-pulses, $\epsilon_{\rm{R}}$ of mapping ramps, and $\epsilon_{\rm{M}}$ of the measurement point. Exchange energy, $J_{\rm{E}}$ used for $\pi$ pulses is indicated. (d) Bloch sphere of the singlet-triplet qubit, with mechanisms of rotation indicated, exchange energy, $J$, and Zeeman field difference along the applied field direction, $\Delta B_{\rm{z}}$, drive rotations around the qubit $z$-axis and $x$-axis respectively. 
}
\end{figure}

In this Letter, we demonstrate echo recovery of singlet-triplet superposition amplitude by embedding qubit rotations about two axes of the Bloch sphere between CP sequences. For Overhauser-driven rotations about the $x$ axis of the Bloch sphere, and exchange-driven rotations about the $z$ axis, ensembles of single-shot singlet measurements show coherent oscillations for total sequence times (including echoes) of $60~\mu$s. We then compare singlet recovery using Hahn, CP, concatenated dynamical decoupling (CDD) \cite{Khodjasteh, WitzelCDD2007}, and Uhrig dynamical decoupling (UDD) \cite{Uhrig2007} schemes. We find that for the present setup, a 16-pulse CP sequence and 21-pulse (fifth order) CDD sequence yield comparable performance, both with $T_2\sim 80~\mu$s, while a 22-pulse (22nd order) UDD performs less well, possibly due to the limited bandwidth of the pulses.

The double quantum dot and sensor are defined by Ti/Au depletion gates on a GaAs/Al$_{0.3}$Ga$_{0.7}$As heterostructure with a two-dimensional electron gas  (density $2\times10^{15} ~\rm{m}^{-2}$, mobility $20~\rm{m}^2$/Vs) $100$ nm below the surface. An in-plane magnetic field, $B=750$~mT, was applied perpendicular to the dot connection axis, as indicated in Fig.~\ref{T2Fig1}(a). Measurements were carried out in a dilution refrigerator (electron temperature $\sim 150$~mK) configured for high-bandwidth gating and rf reflectometry.  The state of the double quantum dot was controlled by pulsed gate voltages $V_{\rm{L}}$, $V_{\rm{R}}$ [Figs.~\ref{T2Fig1}(a,b)] using a Tektronix AWG 5014. Except for point P, where the (0,2) singlet is prepared, gate configurations fall on a line between (0,2) and (1,1) charge states, parameterized by the detuning, $\epsilon$ [Fig.~\ref{T2Fig1}(b)].  To reduce effects of voltage drift, average gate voltages were set to the separation point S using a compensation pulse between measurement and preparation \cite{BluhmT2}. The charge state of the double dot was detected using rf reflectometry \cite{Reillyapl07} of a proximal sensor quantum dot, integrated over $\sim600~$ns to yield a single-shot measurement, as described previously \cite{Singleshotpaper,Dotsensorpaper}.

All pulse sequences start by allowing two electrons at point P to relax to the (0,2) singlet [Fig.~1(c)]. Fast separation of the electrons to point S initializes the system into the $(1,1)$ singlet, the $+z$ direction on the qubit Bloch sphere [Fig.~1(d)]. Alternatively, fast separation from P to R followed by adiabatic separation from R to S initializes the system into the (1,1) ground state of the Overhauser fields,  $\mid\uparrow\downarrow\rangle$, the $+x$ direction on the Bloch sphere~\cite{petta05}. At point S, exchange splitting, $J_{\rm{S}}$, is negligible but Overhauser gradients are not, and the qubit precesses rapidly around its $x$ axis at frequency $f_{\rm{S}}=g^* \mu_{\rm{B}} \Delta B_{\rm{z}}/h$, where $g^* \sim -0.4$ is the GaAs g-factor and $\Delta B_{\rm{z}}$ is difference in Zeeman fields (in this case, Overhauser fields) along the direction of applied field. Controlled rotation about the $z$-axis of the Bloch sphere is realized by pulsing from S to E, where a large exchange energy, $J_{\rm{E}}$, causes qubit rotation at frequency $f_{\rm{E}} = \sqrt{J_{\rm{E}}^2+(g^*\mu_{\rm{B}} \Delta B_{\rm{z}})^2}/h$. Frequency $f_{\rm{E}}$ includes a small quadrature contribution from $\Delta B_{\rm{z}}$ that causes inhomogeneous dephasing and alters the rotation axis during the exchange pulse [Fig.~1(d)].   After evolution at S and E, the qubit is measured by pulsing to the measurement point, M. The singlet state can recombine in (0,2) while the triplet state remains in $(1,1)$. The resulting charge state difference is detected by the sensor quantum dot as $V_{\rm{rf}}$ \cite{Singleshotpaper,Dotsensorpaper}.

  \begin{figure}[t]
\includegraphics{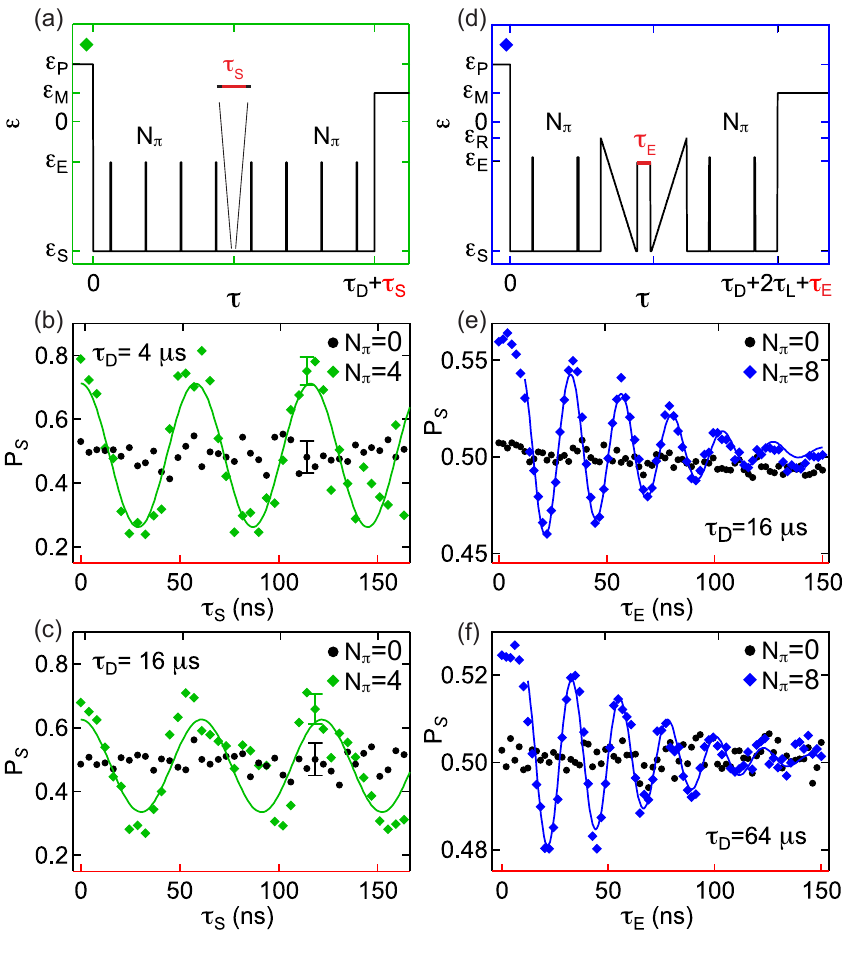}
\caption{
\label{T2xrot}
\label{T2zrot}(Color online) (a) Pulse sequence for $x$ rotations, cycling through preparation (P), separation (S), exchange-echo (E), and measurement (M) points in gate-voltage space. Evolution at S for time  $\tau_{\rm S}$ produces $x$ rotations due to Overhauser gradient. (b) Singlet return probability, $P_S$, as function of $\tau_{\rm{S}}$ for $N_{\pi}=4$ (green diamonds) and $N_{\pi}=0$ (black circles), with $\tau_{\rm{D}}=4~\mu$s, along with cosine fit (curve) to data, post-selected to have a specific period (see text). Without echo pulses ($N_{\pi}=0$), no oscillations are seen. Error bar reflects statistics of $100$ single-shot measurements per point. (c) Identical to (b) with $\tau_{\rm{D}}=16~\mu$s, showing reduced visibility. (d) Pulse sequence for $z$ rotations: after an initial CP sequence, singlet is mapped to $\mid\uparrow\downarrow\rangle$ by pulsing to R and ramping to S. Pulsing to E for $\tau_{\rm{E}}$ induces rotation around $z$. Pulsing to S then ramping to R then pulsing back to S maps $\mid\uparrow\downarrow\rangle$ back into the singlet. After a second CP series, pulsing to M yields single-shot measurement. (e) Singlet probability, $P_S$, as function of $\tau_{\rm{S}}$ for $N_{\pi}=8$ (blue diamonds) and $N_{\pi}=0$ (black circles), with $\tau_{\rm{D}}=16~\mu$s,  based on an average of $600\times100$ single-shot measurements, along with cosine fit with gaussian envelope, see text. Error bars from statistics are smaller than markers. (f) Identical to (e) with $\tau_{\rm{D}}=64~\mu$s.
}
\end{figure}

Dephasing of the $S$-$T_0$ qubit can arise from thermally driven evolution of the Overhauser field difference between dots or from electron motion in a spatially varying Overhauser field caused by gate noise or drift. Here, we do not distinguish between decoherence due to quantum entanglement with the environment and dephasing due to a noisy classical environment. Because Overhauser fluctuations are concentrated below 1 Hz ~\cite{ReillyCorr08}, apparent dephasing due to precession of the qubit can be readily recovered using dynamical decoupling schemes applied on faster time scales~\cite{petta05,BluhmT2}. The simplest such scheme, Hahn echo (HE), uses a single $\pi$-pulse, realized in this case by pulsing to E for a time $\pi/(\hbar J_E)$ after a (dephasing) interval $\tau_{\rm{D}}/2$ at S, followed by a second (rephasing) interval at S for $\tau_{\rm{D}}/2$. Multiple $\pi$-pulses applied within a given interval can extend coherence by shortening the interval between dephasing and rephasing, during which the environment may have evolved \cite{CP}. 
The simplest multi-pulse scheme, the CP sequence, inserts several of equally spaced $\pi$-pulses surrounded by dephasing and rephasing intervals \cite{CP, CPisCPMG}.

As a first approach to dynamical decoupling of arbitrary qubit states, we investigate $x$ and $z$ rotations interlaced between two CP sequences. For this purpose, the sequence preceding the rotations is not needed, but including it serves to demonstrate general interlacing of qubit operations with dynamical decoupling. 

To interlace $x$ rotations between CP sequences [see Fig.~\ref{T2xrot}(a)], the (1,1) singlet was first initalized, followed by a $N_\pi = 4$ CP sequence during an interval $\tau_{\rm{D}}/2$. Pausing at S for a time $\tau_{\rm{S}}$ induced rotation around $x$ axis due to the slowly-varying Overhauser field gradient.  This was followed by a second $N_\pi = 4$ CP sequence, and finally a single-shot measurement at point M.  
For each value of $\tau_{\rm{S}}$, this sequence was repeated 100 times, yielding a probability, $P_S$, of a singlet outcome over these 100 single-shot events. Probabilities for 40 values of $\tau_{\rm{S}}$, ranging from 0 to 180 ns, were measured over a total duration of 100 ms. Oscillations of $P_S(\tau_{\rm{S}})$ are well fit by $P_S(\tau_{\rm{S}}) =0.5\,\left(1+V \cos{\left(\tau_{\rm{S}}g^*\mu_{\rm{B}}\Delta B_{\rm{z}}/\hbar\right)}\right)$, where $V$ is the measurement visibility [Figs.~2(b,c)], the expected form for qubit precession in a fixed Overhauser field difference $\Delta B_{\rm{z}}$~\cite{Singleshotpaper}. Averaging $x$ rotations over the full range of Overhauser fields would wash out oscillations. Instead, $P_S(\tau_{\rm{S}})$ data in Figs.~2(b,c) was selected to cover a narrow range of Overhauser fields around $\Delta B_{\rm{z}}\sim 3~$mT. The value of $\Delta B_{\rm{z}}$ evolves slowly among data sets, observable as a fluctuating oscillation period of $P_S(\tau_{\rm{S}})$.

 From the ratio of visibilities for long and short (but otherwise identical) sequences, $V(\tau_D = 16 \mu {\rm s})/V(\tau_D = 4 \mu {\rm s}) = 0.88$, we estimate a dephasing time $T_2\sim 40\,\mu$s  for the complete sequence with $N_\pi = 4$ before and after the rotation, assuming a gaussian decay envelope (discussed below) and using the measured normalizations $V_0\sim 0.70$ with $\tau_{\rm{D}}=0$, $N_\pi = 0$, and $V_0\sim 0.63$ with $\tau_{\rm{D}}= 0$,  $N_\pi = 4$. The reduced visibility of $S$-$T_0$ precession without the CP sequences can be attributed to limited read-out fidelity, finite exchange at $\epsilon_{\rm{S}}$, and imperfect initialization in (1,1). A further $\sim 10\%$ reduction in visibility when including $2N_\pi = 8$ pulses is dominated by tilting of the echo rotation axis away from $z$ by Overhauser field gradients [Fig.~\ref{T2Fig1}(d)]. 

\begin{figure}[t]
\centering
\includegraphics{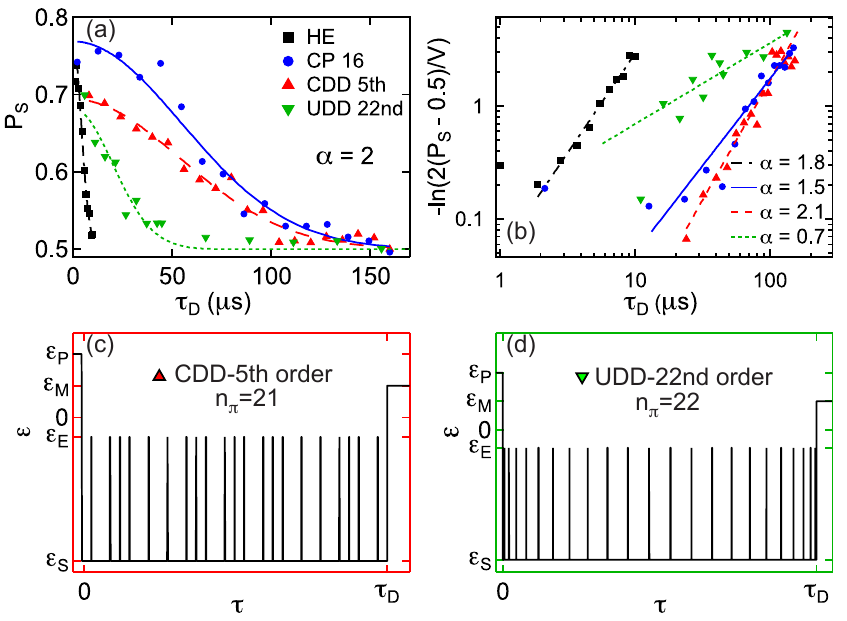}
\caption{
\label{T2decouplings}
(Color online) (a) Singlet recovery amplitude, $P_S$, as a function of total dephasing time, $\tau_{\rm{D}}$, for Hahn, CP, CDD and UDD sequences (see text), with fits of Eq.~(\ref{Envelope}) with $\alpha=2$, yielding $T_2^{\rm{HE}}\sim6~\mu$s,  $T_2^{\rm{CDD}}\sim80~\mu$s,  $T_2^{\rm{CP}}\sim80~\mu$s,  $T_2^{\rm{UDD}}\sim30~\mu$s~\cite{T2params}. (b) Replotting data in (a) as $-{\rm ln}[2(P_S - 1/2)/V]$ allows a fit giving the exponent $\alpha$ in Eq.~(1). On a log-log plot, Eq.~(1) and data appears linear with slope $\alpha$ \cite{T2params}. Pulse sequences for (c) $5$th order CDD and (d) $22$nd order UDD. 
}
\end{figure}

Interlaced $z$ rotations between CP sequences were demonstrated using adiabatic ramps that map $S$ and $T_0$ to $\mid\uparrow\downarrow\rangle$ and $\mid\downarrow\uparrow\rangle$ [Fig.~\ref{T2zrot}(d)] \cite{petta05}. Initialization of the (1,1) singlet was followed by a $N_\pi = 8$ CP sequence. The preserved singlet was then mapped to $\mid\uparrow\downarrow\rangle$ by pulsing from S to R (without passing the $S$-$T_+$ anticrossing) then ramping slowly to S in a (loading) time $\tau_{\rm{L}}\sim400$ns. Pulsing to E for a time $\tau_{\rm{E}}$ induced rotation around $z$ at frequency $f_{\rm{E}}$. Next, pulsing to S, ramping slowly from S to R, then back to S mapped  $\mid\uparrow\downarrow\rangle$($\mid\downarrow\uparrow\rangle$) to $S$($T_0$).  The resulting superposition amplitude was preserved over a time $\tau_{\rm{D}}/2$ by a second $N_\pi = 8$ CP sequence.  Finally, the superposition amplitude was measured by pulsing to M. 600 singlet probabilities, each based on 100 single-shot measurements, were averaged over fluctuating Overhauser fields for each $\tau_{\rm{E}}$. Oscillations in $P_S(\tau_{\rm{E}})$ do survive averaging over Overhauser fields, since for $J_{\rm{E}}\gg\Delta B_{\rm{z}}$ the oscillation frequency $f_{\rm{E}}$ depends only weakly on $\Delta B_{\rm{z}}$. However, this averaging does contribute to the decay envelope seen in Figs.~2(e,f).  Note that mapping between $z$ and $x$ axes accurately preserves the angle of the superposition from the corresponding axis (amplitude), with some loss of azimuthal angle (phase) information due to environment dynamics on the time scale of the 400 ns ramp. State tomography \cite{Foletti09} requires fast rotations compared to environment dynamics \cite{ReillyCorr08}. 

Singlet return probability $P_S(\tau_{\rm{E}})$, averaged over fluctuating Overhauser fields, for $\tau_{\rm{D}}=16~\mu$s [Fig.~2(e)] and $\tau_{\rm{D}}=64~\mu$s [Fig.~2(f)] using $N_\pi = 8$ CP sequences show damped oscillations a function of the exchange time $\tau_{\rm{E}}$. Fits to $P_S =1/2+(V/2) \cos{\left(\tau_{\rm{E}} J_{\rm{E}}/\hbar\right)}\exp [{-(\tau_{\rm{E}}/T_2^*)^2}]$, the expected form for a qubit precessing in exchange splitting, $J_{\rm{E}}$, with inhomogenous dephasing due to both fluctuating Overhauser fields and electrical noise, yield $J_{\rm{E}}\sim0.2~\mu$eV  and $T_2^*\sim100$~ns from the decay envelope~\cite{zrotparam}. Visibilities $V\sim0.09 (0.06)$ for $\tau_{\rm{D}}=16(64)~\mu$s, along with $V \sim0.3$ measured without CP sequences (not shown), yield a coherence time $T_2 \sim 80~\mu$s. Besides finite coherence time ($T_2$), other contributions that reduce visibility include nonzero $\Delta B_{\rm{z}}$ during exchange pulses, nonzero exchange at the separation point, $J_{\rm{S}}$, and gate noise during $\pi$-pulses. 

Next, we consider alternative dynamical decoupling schemes for recovery of a prepared singlet without interlaced rotations. A prepared (1,1) singlet at point S was given $n_\pi$ $\pi$-pulses, spaced in time according to the particular decoupling scheme, over a total interval $\tau_{\rm{D}}$. 
For each value of $\tau_{\rm{D}}$, an Overhauser-averaged singlet return probability $P_S$ was found by averaging $\sim10^4$ single-shot measurements, normalized by reflectometer output voltages of singlet and triplet outcomes via single-shot histograms~\cite{Singleshotpaper}. For all decoupling schemes, $P_S(\tau_{\rm D})$ data are well described by the functional form

\begin{equation}
\label{Envelope}
P_S(\tau_{\rm D}) = 1/2 + (V/2)\,e^{-(\tau_{\rm{D}}/T_2)^{\alpha}}.
\end{equation}
We either set $\alpha = 2$ [Fig.~3(a)] or leave $\alpha$ as a fit parameter [Fig.~3(b)]. Allowing $\alpha$ to vary can give insight into the dominant dephasing mechanism, depending on the decoupling scheme \cite{WitzelCP,YaoT2}. For instance, for simple Hahn echo, a white-noise environment is expected to give exponential decay, $\alpha =1$. In contrast, the enhanced low-frequency content of an Overhauser-field-dominated environment is expected to yield $\alpha \sim 4$ for Hahn echo \cite{WitzelCP,YaoT2}, as observed experimentally~\cite{BluhmT2}.

Experimental $P_S(\tau_{\rm D})$, for several decoupling schemes along with fits of Eq.~(\ref{Envelope}) with fixed $\alpha=2$ yields values for $T_2$ [Fig.~3(a)].  For a single-pulse Hahn echo, $T_2^{\rm{HE}}\sim6~\mu$s, while $T_2^{\rm{CP}}\sim80~\mu$s for a CP sequence with $16$ $\pi$-pulses~\cite{T2params}. A comparison of CP sequences with different $N_\pi$ are shown in Fig.~3(b). Best-fit values for $\alpha$ are extracted by taking the logarithm of Eq.~(1), as shown in Fig.~\ref{T2decouplings}(c). For Hahn echo, the best fit value is $\alpha^{\rm{HE}} = 1.8$, suggesting that electrical (gate) noise and drift, combined with spatially varying Overhauser fields, are likely the dominant source of dephasing, rather than time dependence of the Overhauser fields themselves. For the CP sequence, the best fit exponent is $\alpha^{\rm{CP}}\sim1.5$. 

Alternative decoupling sequences may outperform CP, depending on the decohering environment \cite{WitzelCDD2007,Uhrig2007, Lee2008}. A favorable scheme for spin qubits coupled to nuclear environments is CDD whose $n$th order sequence is created from two sequences of ${(n-1)}$th order with an additional $\pi$-pulse between them for odd $n$~\cite{WitzelCDD2007}: first-order CDD is Hahn echo; second-order CDD is CP with two $\pi$-pulses; fifth-order CDD, with $21$ $\pi$-pulses, is shown in Fig.~\ref{T2decouplings}(c). Another pulse scheme optimized for spin-bath environments is UDD, which has $n_\pi$ $\pi$-pulses (indexed by $j$) at times $\delta \tau = \tau_{\rm{D}}\,\sin^2[j\pi/(2n+2)] $.  The $22$nd-order UDD ($n_\pi$ = 22) is shown in Fig~\ref{T2decouplings}(d). Singlet probabilities $P_S(\tau_{\rm D})$ for the $5$th-order CDD and $22$nd-order UDD sequence is shown in Fig.~\ref{T2decouplings}(a) along with Hahn and CP, and yield $T_2^{\rm{CDD}} \sim 80~\mu$s and $T_2^{\rm{UDD}}\sim30~\mu$s~\cite{T2params}. We note that (i) all multi-pulse sequences significantly outperform Hahn echo; (ii) the fifth-order CDD sequence ($n_\pi = 21$) has no better performance than the $n_\pi = 16$ CP sequence, and (iii) the $T_2$ achieved for UDD is considerably shorter than for comparable CDD and CP sequences. This presumably results from experimental artifacts such as pulse bandwidth limitations, but could also reflect the non-optimality of UDD for an environment with a $1/{f^2}$ high-frequency tail \cite{ReillyCorr08}. For a dephasing power spectrum closer to the experiment the optimal spin echo pulse sequence has been predicted to be very similar to a simple CP sequence~\cite{Pasini2010}, while in Ref.~\onlinecite{Lee2008} it was predicted that UDD is optimal for hyperfine-induced dephasing in GaAs quantum dots. Optimizing pulse sequences with contributions from gate noise and drift combine with spatially and temporally varying Overhauser fields remains an outstanding problem experimentally and theoretically. \\
\begin{acknowledgments}
CB and JM contributed equally to this work. We acknowledge funding from iARPA/ARO and the Department of Defense.  We thank Hendrik Bluhm, Sandra Foletti, Daniel Loss, David Reilly, and Wayne Witzel for useful discussion.
\end{acknowledgments}

\appendix


\begin{thebibliography}{10}
\bibitem{zrotparam} An offset of $\sim 10~$ns due to finite pulse-rise time is taken into account for the fit.
\end{thebibliography}

\end{document}